\documentclass[conference]{IEEEtran}
\IEEEoverridecommandlockouts
\usepackage{cite}
\usepackage{amsmath,amssymb,amsfonts}
\usepackage{algorithmic}
\usepackage{graphicx}
\usepackage{textcomp}
\usepackage{makecell} 
\usepackage{balance}
\usepackage{subcaption}
\usepackage{hyperref}
\usepackage{caption}
\usepackage{float}
\usepackage{url}
\usepackage{xcolor}
\def\BibTeX{{\rm B\kern-.05em{\sc i\kern-.025em b}\kern-.08em
    T\kern-.1667em\lower.7ex\hbox{E}\kern-.125emX}}
\begin{document}

\title{\taskname: A Framework to Simulate Self-Reported Ground Truth for Mental Health Sensing Studies \\
}
\author{\IEEEauthorblockN{Akshat Choube}
\IEEEauthorblockA{ 
\textit{Northeastern University}\\
Boston, Massachusetts, USA \\
choube.a@northeastern.edu}
\and
\IEEEauthorblockN{Vedant Das Swain}
\IEEEauthorblockA{
\textit{Northeastern University}\\
Boston, Massachusetts, USA \\
v.dasswain@northeastern.edu}
\and
\IEEEauthorblockN{Varun Mishra}
\IEEEauthorblockA{\textit{Northeastern University} \\
Boston, Massachusetts, USA \\
v.mishra@northeastern.edu}
}
\newcommand{\taskname}[1]{\textbf{SeSaMe}#1}
\newcommand{\tasknamescale}[1]{\textbf{SeSaMe$_{\text{scale}}$}#1}

\maketitle

\begin{abstract}


Advances in mobile and wearable technologies have enabled the potential to passively monitor and track a person's behavior for various mental, behavioral, and affective health outcomes. 
These approaches typically rely on longitudinal collection of self-reported outcomes, e.g., depression, stress, and anxiety, to 
train machine learning models. 
However, the need to continuously self-report various internal states adds a significant burden on the participants, often resulting in attrition, missing labels, or insincere responses.
In this work, we introduce the Scale Scores Simulation using Mental Models (SeSaMe) framework to alleviate participants' burden in digital mental health studies. By leveraging pre-trained large language models (LLMs), \taskname\ enables the simulation of participants' responses on psychological scales.
In \taskname, researchers can prompt LLMs with information on participants' internal behavioral dispositions, enabling LLMs to construct mental models of participants to simulate their responses on psychological scales.
As part of the framework, we provide four evaluation metrics to assess the effectiveness of the simulated responses. 
We demonstrate an application of \taskname, where we use GPT-4 to simulate responses on one scale using responses from another as behavioral information. 
We use \taskname's evaluation metrics to assess the alignment between human and \taskname-simulated responses to psychological scales. Then, we present multiple experiments to inspect the utility of \taskname-simulated responses as ground truth in training machine-learning models by replicating established depression and anxiety screening tasks with passive sensing data from a previous study.
Our results indicate \taskname\ to be a promising approach, but its alignment may vary across scales and specific prediction objectives. We also observed that model performance with simulated data was on par with using the real data for training in most evaluation scenarios.  
We conclude by discussing the potential implications of \taskname\ in addressing some challenges researchers face with ground-truth collection in passive sensing studies.

\end{abstract}


\section{Introduction}
Mental health continues to be a pressing concern worldwide. In the United States, more than one out of five adults struggles with a mental illness \cite{nimh-mental-illness}, and challenges posed by the pandemic, economic uncertainties, and rapid social changes have intensified these struggles \cite{lee2021impact}. Consequently, traditional mental health screening and care systems relying on in-person interviews and self-reported experiences face considerable strain in scaling to the growing population with mental health needs~\cite{cosic2020impact}. Digital mental health screening methods, which use passively gathered data from mobile phones, wearables, and social media, offer a promising alternative \cite{mohr2017personal}. These sensing technologies, being non-intrusive and fully automated, promise greater scalability \cite{abdullah2018sensing}.

Over the past decade, a multitude of passive sensing technologies have employed supervised machine learning (ML) and deep learning methods to assess various mental health markers, including depression \cite{xu2023globem, tlachac2022deprest, de2013predicting, wang2014studentlife}, anxiety \cite{tlachac2022deprest, arif2020classification, fukazawa2019predicting, saha2021person}, and stress \cite{mishra:context, sano2013stress, mishra:continuous-stress, mishra:stress-reproducibility, DasSwain2022semantic}. To train supervised models, ground truth labels are necessary alongside passively sensed data; the standard practice is to employ questionnaires and Ecological Momentary Assessments (EMAs) to capture the ground truth. Researchers extensively use clinically-validated psychological scales (like Patient Health Questionnaire-9 (PHQ-9) \cite{kroenke2001phq}, Generalized Anxiety Disorder-7 \cite {spitzer2006brief}, etc.)  to gauge the presence and severity of participant's mental health conditions serving as the ground truth. Researchers often require participants to fill out multiple questionnaires to measure various mental health conditions, which increases the burden on participants, leading to incomplete or insincere responses \cite{huang2015insufficient, campbell:PatientEngagementMultimodal-2023}. This effect is more pronounced in longitudinal studies where participants are required to fill out surveys on a weekly or daily basis over an extended period \cite{vhaduri2017design}. This is a major challenge that needs to be addressed to fully harness the potential of digital health systems.


In the past, researchers have attempted to make the questionnaire interfaces more engaging by using chatbots \cite{filler2015mobilecoach, welch2020using}, large language models (LLMs) \cite {wei2023leveraging, yun2023keeping}, virtual agents \cite{pfeifer2011longitudinal, vardoulakis2013social}. These approaches, however, still demand significant time commitment and are considered burdensome~\cite{chan2018students}. Prior works have also explored building ML models to predict participants' responses on questionnaires using their responses on other questionnaires \cite{jivnani2023predicting} or social media posts \cite{vu2020predicting}. Building ML models from scratch requires considerable time and effort. Furthermore, integrating new data sources requires modifying model architecture and retraining.

We investigate an alternate approach to ease the participants' burden of responding to multiple psychological scales, by simulating their responses using LLMs. Recent research in pre-trained LLMs (GPT-4 \cite{openai2023gpt}, Llama \cite{touvron2023llama}, and Gemini \cite{team2023gemini}) has demonstrated their capabilities to perform well in zero-shot settings (without additional data and training)\cite{kojima2022large, wei2021finetuned}, offering quick and low effort off-the-shelf solutions to problems in various domains including mental health \cite{xu2024mental}. Additionally, LLMs have shown the ability to mimic human behavior, prompting researchers across disciplines to use them to simulate human behavior in experiments \cite{argyle2023out, aher2023using, hamalainen2023evaluating}. In this work, we propose a new framework called  \textbf{S}cal\textbf{e} \textbf{S}cores Simul\textbf{a}tion using \textbf{Me}ntal Models (\taskname) that leverages information on the participants' internal behavioral dispositions to enable LLMs to build mental models of participants to simulate their responses on a psychological scale. We define four evaluation metrics to measure the alignment of responses simulated using this framework with the participants' real data. We then demonstrate an application of this framework, using participants' responses to a different scale as behavioral information used to simulate their responses on the desired scale. For example, we simulate a participant's responses to GAD-7 items by prompting an LLM with the participant's scores on PHQ-9 items. 

As scores on these psychological scales serve as ground truth in training mental health sensing models, it was imperative to assess ML models trained on simulated scores. To this end, we replicated a prior mental health study on anxiety and depression screening \cite{tlachac2022deprest} using LLM-simulated PHQ-9 and GAD-7 scores as ground truth for training ML models and conducted multiple experiments to analyze its impact.

We performed all our experiments using  GPT-4 as LLM. Our results show that GPT-4 exhibited subpar performances across various simulation evaluation criteria, with performance disparities observed depending on the scale generated and the scale used as behavioral information. However, GPT-4 demonstrated proficiency in simulating PHQ-9 with GAD-7 and vice versa. The models trained on simulated PHQ-9 and GAD-7 scores as ground truth outperformed the original models in anxiety screening but exhibited comparable performance in mild depression screening and inferior performance in moderate to high depression screening. 
Our code and results are publicly available at \href{https://github.com/UbiWell/sesame-llms-ground-truth-code}{https://github.com/UbiWell/sesame-llms-ground-truth-code}.

Our work makes the following contributions:

\begin{itemize}
    \item We introduce a novel framework called \taskname\  to simulate participants' responses on psychological scales, present four simulation evaluation metrics, and then show an application of this framework using participants' responses on a different scale as behavioral information.
    \item We explore the impact of participants' scale scores simulated using \taskname\  by replicating a previous digital health sensing experiment on anxiety and depression screening \cite{tlachac2022deprest}, using simulated scores as the ground truth for ML model training.
    \item We delve into the potential implications of using \taskname\ simulated responses in research and development. 
    
\end{itemize}

    

\section{Related Work}
\subsection{Digital Mental Health Sensing}
Traditional clinical approaches for detecting mental health disorders are intrusive and lack real-time monitoring capabilities. Our interactions with technology yield significant behavioral data, making digital sensing methods a powerful and promising enabler for real-time and unobtrusive mental health sensing \cite{abdullah2018sensing}. Studies focusing on inferring human behavioral and mental health outcomes require participants to complete questionnaires, using their responses as the ground truth for modeling. Wang et al. \cite{wang2014studentlife} required participants to fill Perceived Stress Scale (PSS)\cite{cohen1994perceived}, UCLA Loneliness Scale \cite{russell1996ucla}, Flourishing Scale \cite{diener2010new}, and PHQ-9 questionnaires to gauge different aspects of mental well-being. Likewise, Xu et al. \cite{xu2023globem} leveraged passive mobile sensing data to construct machine learning models for depression detection, with participants completing the PHQ-4 weekly, serving as the ground truth. Additionally, participants were tasked with completing other questionnaires, such as PSS and PANAS \cite{watson1988development}, on a weekly basis, along with several more questionnaires administered both before and after the study. Tlachac et al. \cite{tlachac2022deprest} used call and text logs to detect anxiety and depression from GAD-7 and PHQ-9 scale responses from participants as ground truth. While these questionnaires aid in streamlining mental health screening, they are frequently considered burdensome by the participants, resulting in low completion rates and high attrition \cite{young2006attrition}. This phenomenon is particularly prevalent in longitudinal studies where participants are requested to complete weekly or daily questionnaires over an extended duration \cite{moller2013investigating, vhaduri2017design}. Additionally, self-reporting mental health conditions is also subject to stigma \cite{gulliver2010perceived, DasSwain2022semantic}, and individuals grappling with mental health issues are less responsive to these questionnaires \cite{proudfoot2013impact, brueton2014use}. Thus, numerous obstacles contribute to incomplete or insincere responses to mental health questionnaires. Recent advancements in LLMs that empower them to replicate human behavior offer a potential approach to simulate these responses.

\subsection{Simulating Humans using Large Language Models (LLMs)}

LLMs \cite{radford2019language, brown2020language, openai2023gpt} have shown exceptional capabilities to mimic human-like behavior and are being employed to create lifelike agents posing as personal assistants \cite{pi_ai_talk}, tutors \cite{pardos2023learning}, and therapists \cite{wang2021evaluation}. Consequently, an intriguing question posed to researchers in the fields of Human-Computer Interaction, Social Science, and Psychology is whether LLMs can serve as partial or even complete replacements for human participants in experiments. Argyle et al. \cite{argyle2023out} demonstrated that GPT-3 models can emulate a range of realistic and diverse political opinions when provided with the background stories of individuals. Aher et al. \cite{aher2023using}, using LLMs, replicated established economics, psycholinguistics, and psychology experiments, demonstrating similar trends as seen with real participants. Tavast et al. \cite{tavast2022language}, evaluated the diversity and human-likeliness of synthetic self-report data generated by different GPT-3 model variants. 
In their subsequent work \cite{hamalainen2023evaluating}, the authors tested the capabilities of GPT-3 models in generating responses to open-ended interview questions. Their findings indicated that LLMs are capable of generating data comparable to that of humans, although the generated data tends to be less diverse. Researchers have also built mini-worlds with multiple LLM agents simulating humans \cite{park2023generative, liu2023training}.
Some companies like syntheticusers.com \cite{syntheticusers} are building infrastructures to conduct experiments using synthetic AI participants.

In the midst of research demonstrating LLMs as potential substitutes for humans, there are multiple studies advocating caution before employing LLMs as human equivalents in experiments.  Ullman \cite{ullman2023large} demonstrated that LLMs fail on trivial modifications to theory-of-mind tasks. Researchers have argued that training methods and data on which LLMs are trained impede them from being valid proxies to humans \cite{wang2024large, tjuatja2023llms}. The extent to which LLMs can accurately simulate human behavior remains a topic of considerable debate within the research community \cite{dillion2023can, harding2023ai}.


Nevertheless, LLMs are transitioning beyond being viewed solely as tools to assuming more anthropomorphic forms. This paradigm shift has sparked researchers' curiosity in understanding LLMs' inherent traits by instructing them to respond to psychometric scales birthing the field of AI Psychometry.


\subsection{AI Psychometry}
Researcher have prompted LLMs with different psychometric scales to analyze their psychological profiles. La et. al \cite{la2024open} conducted a series of assessments on LLMs
to evaluate their inherent personality traits and investigate how assigning them roles influences their intrinsic personality. Their results showed that different LLMs have different intrinsic personalities, and assigning them roles had varying effects, with most models retaining their intrinsic personalities. Conversely, \cite{safdari2023personality} showed that LLMs outputs can be shaped reliably to mimic human personalities. Similarly, \cite{huang2023chatgpt} built a framework called "PsychoBench" and evaluated various LLMs' responses on 13 psychometric scales, concluding different LLMs inherently have different psychological traits.  

\subsection{Novelty}

Previous studies have prompted LLMs to respond to psychological scales either to simulate group trends in human behavior or to test their inherent psychological profiles. However, simulating a specific participant's responses to psychological scales based on an LLM's mental model of the participant has not been explored yet. We not only introduce a framework for this purpose but also demonstrate its application. Furthermore, we present an unprecedented evaluation of using simulated responses as ground truth for ML model training in a mental health sensing study.



\section{\textbf{S}cal\textbf{e} \textbf{S}cores Simul\textbf{a}tion using \textbf{Me}ntal Models (\taskname)}

We present a novel framework for simulating participants' itemized scores on a scale ($S_A$) using LLMs based on their mental model of participants (\taskname). LLMs, trained on extensive datasets, excel in simulating group trends in human behavior. To simulate a specific participant, however, we need to prompt LLMs with information ($I$) on that participant's behavioral dispositions. We formally define this as:

$$ \hat{R}_A = LLM(I, S_A)$$
$\hat{R}_A$ is the itemized scores to scale $S_A$ simulated by an LLM. 

We can apply this framework to simulate responses $[\hat{R}_A]$ for all the participants ($n$) in our experiment. To evaluate the simulated scores, we require the real scores of the participants $[R_A]$. Psychological scales typically employ a heuristic to combine scores on individual scale items into an aggregated score, used for screening specific psychological constructs. By applying this heuristic, we can obtain aggregated real scores ([$T_A$]) and aggregated simulated scores ([$\hat{T}_A$]).

We define the following four metrics for evaluating the alignment of simulated scores with real scores:

\begin{itemize}
\setlength{\itemsep}{8pt}
 \item \textbf{Simulation Error}: The simulation error can be measured at the itemized scores or aggregated scores level.
  $$e_{aggregated} = \frac{\sum_{i=1}^{n} | T_{A}^{i} - \hat{T}_{A}^{i}|}{n*max\_score(S_A)}$$ 
  $$e_{itemized} = \frac{\sum_{i=1}^{n} \sum_{j=1}^{|S_A|} |R_{A_j}^{i} - \hat{R}_{A_j}^{i}|}{n*max\_score(S_A)}$$ 
  
 Here $max\_score(S_A)$ and $|S_A|$ denote the maximum score possible and number of items on scale $S_A$, respectively. An itemized simulation error of zero implies hyper-realistic simulation.

\item \textbf{Correlation Alignment}:
The correlation between aggregated real and simulated  scores ($corr([T_A], [\hat{T}_A])$) should be strong ($r > 0.7$), where $corr(.)$ denotes the correlation metric used.

\item \textbf{Distribution Alignment}: The density distributions of $[\hat{T}_A]$ and $[T_A]$ should be similar.

\item \textbf {Cross-Scale Validity}: During studies, participants often complete multiple psychological scales measuring different mental health constructs. The simulated responses on a scale should be consistent with participants' responses on other scales. Given participants' aggregated scores $[T_B]$ on another scale $S_B$, we can measure cross-scale validity as:
 \begin{itemize}
    \item {Trend Alignment}: \\$sgn(corr([T_B], [T_A])) = sgn (corr([T_B], [\hat{T}_A]))$ \\ where $sgn(.)$ denotes the sign function.

    \item {Correlation Strength}:\\ $corr([T_B], [T_A]) \approx corr([T_B], [\hat{T}_A])$.
    
\end{itemize}

\end{itemize}

Note that some of these metrics are necessary but not sufficient for a successful simulation. For example, high distribution alignment is not sufficient as individual scores may deviate significantly from real scores despite similarities in overall distribution across participants. Thus, we need to look at collective results of these metrics to better interpret the performance of simulation.


\section{Datasets and Method}
In this section, we present an application of \taskname\ framework where we simulated participants' responses on a scale using their responses to a different scale as behavioral information provided to LLMs. Formally, given $R_A$ and $R_B$ as real itemized scores of person $X$ on scale $S_A$ and $S_B$, respectively, we simulated $X$'s itemized scores $\hat{R}_A$ on $S_A$, using their real responses $R_B$ on scale $S_B$ as information for LLMs to build a mental model of $X$ ($I = (S_B, R_B))$. 

We used participants' responses to psychological scales in the following two datasets for our simulations:

\begin{itemize}
  \setlength{\itemsep}{5pt} 
   \item \textbf{StudentLife Dataset}: StudentLife Dataset \cite{wang2014studentlife} comprises passive sensing data, EMA data, and pre- and post-study survey responses from a group of 48 college students. The surveys include mental health measures like PHQ-9, Loneliness Scale, PSS, and Flourishing Scale.

   \item \textbf{DepreST-CAT Dataset}: The DepreST Call and Text (DepreST-CAT) \cite{tlachac2022deprest} dataset consists of participants' call and text logs along with their responses on PHQ-9 and GAD-7 scales. The data is sourced from 369 participants recruited through crowdsourcing.

\end{itemize}


\subsection {\taskname\ using Scores on a Scale as Behavioral Information}





We prompted GPT-4 with person $X$'s scores $R_B$ on scale $S_B$ and instructed it to form an understanding of their mental state. We then instructed it to simulate itemized scores on scale $S_A$ as person $X$ would have. The GPT-4 model simulated itemized scores ($\hat{R}_A$ ) on scale $S_A$. 
We applied scales' scores aggregation heuristics to get $T_B$, $T_A$, and $\hat{T_A}$ for person $X$ based on $R_B$, $R_A$ and $\hat{R}_A$, respectively. We repeated this method for all participants to get a set of aggregated scores for real and GPT-4 simulated scale responses ($[T_B]$, $[T_A]$, and $[\hat{T}_A]$). We then evaluated our simulated responses using the \taskname's evaluation metrics.


For the StudentLife dataset, we used 40 participants' pre- and post-survey responses (a total of 80 responses) on four scales: PSS, PHQ-9, Loneliness Scale, and Flourishing Scale, while for the DepreST-CAT dataset, we used 369 participants' responses on the GAD-7 and PHQ-9 scales. In both datasets, for any pair of scales, we prompted GPT to simulate itemized scores on one scale, given itemized scores on another. We designed the prompt according to Open AI's suggested best practices for prompt engineering \cite{openai_docs}. Figure \ref{fig:prompt} illustrates an example prompt to simulate GAD-7 scores given PHQ-9 scores. We kept default values of hyper-parameters ($temperature=1$ and $top\_p = 1$) in our GPT-4 API calls. We used the Spearman correlation test \cite{spearman1961proof} to assess correlations in our evaluations.

\begin{figure}[h]
\centering
\fbox{\begin{minipage}{1\linewidth}

\fontsize{6.5}{6.5}\selectfont

\medskip

Suppose person 'X' answered these PHQ-9 scale questions from 0 to 3 where 0 denotes 'Not at all', 1 denotes 'Several days', 2 denotes 'more than half days', 3 denotes 'nearly everyday'. Here are the statements:
\\
\\
\hspace{1em} Little interest or pleasure in doing things \\
2 \\
\hspace{1em} Feeling down, depressed, or hopeless \\
1 \\
 . . . \\
\hspace{1em} Thoughts that you would be better off dead, or of hurting yourself \\
1 \\

Based on these responses understand the psychological state of person 'X' and score the given GAD-7 scale questions as the person 'X' would have from 0 to 3. 0 denotes 'Not at all', 1 denotes 'Several days', 2 denotes 'more than half days', 3 denotes 'nearly everyday'. Here are the statements, score them one by one:
\\
\\
\hspace{1em} Feeling nervous, anxious, or on edge \hfill \\
\hspace{1em} Not being able to stop or control worrying \hfill \\
. . . \\
\hspace{1em} Feeling afraid, as if something awful might happen \hfill \\
\\
Just return a list of 7 scores you provided in format [scores] and nothing else

\end{minipage}}
\caption{An example prompt for generating GAD-7 scores from provided PHQ-9 scores. Only a few questions from each scale are represented for brevity, but the actual prompt included all nine PHQ-9 and seven GAD-7 questions.}
\label{fig:prompt}
\end{figure}

\subsection{Usability of \taskname-simulated Labels in Training ML Models in a Mental Health Study}

Further, we wanted to evaluate the utility of \taskname\-simulated scale scores as ground truth in training models in a prior mental health sensing study ``DepreST-CAT" \cite{tlachac2022deprest}. 

In the DepreST-CAT study, the authors developed five separate ML models (Logistic Regression, k-Nearest Neighbour (kNN), Support Vector Machine (SVM), Random Forest (RF), and eXtreme Gradient Boost (XGBoost)) to screen for anxiety and depression from call and text logs, using participants' responses on GAD-7 and PHQ-9  scale as ground truth for model training. 
The authors conducted robust evaluations using leave-one-group-out cross-validation with 100 stratified groups, test data comprising 30\% of the entire dataset. They binarized the training labels based on GAD-7 and PHQ-9 score cutoffs and used the average F1-score across 100 stratified groups as the evaluation metric.

For our experiments, we adopted the same pipeline for feature generation and evaluation; developing the same anxiety and depression screening models as done in the original work but with modifications to the ground truth scores for model training. In \textbf{Experiment B1}, to train anxiety screening models, we completely replaced the real GAD-7 scores ($[R_{GAD}]$) with the GAD-7 scores simulated by GPT-4 ($[\hat{R}_{GAD}]$), using participants' responses on the PHQ-9 scale ($[R_{PHQ}]$). Similarly, to train depression screening models, we replaced $[R_{PHQ}]$ with $[\hat{R}_{PHQ}]$, simulated using $[R_{GAD}]$. To ensure model evaluation on participants' real data, we did not make any alterations to the test data. We compared the performances of depression and anxiety screening models trained with real ground truth scores against those with GPT-4 simulated scores.

We performed three additional experiments to comprehensively evaluate GPT-4 simulated scale scores. 

\begin{itemize}
    \item \textbf{Experiment B2}: To assess the impact of replacing ground truth scores with simulated scores, we did an ablation study, where we increasingly replaced ground truth scores with simulated scores in 30\%, 70\%, and 100\% of training data and compared the model performances.
    
    \item \textbf{Experiment B3}: We compared the performance of models trained on GPT-4 simulated labels with alternative methods, such as Linear Regression and Support Vector Regression. We used 30\% of training data to construct regression models and then replaced real scores with predicted scores for the remaining 70\% of training data.

    \item \textbf{Experiment B4}: We enhanced the original prompts to GPT-4 by incorporating additional demographic information, including age, gender, and a yes/no question on prior depression treatment, collected during the start of the original study. We then evaluated ML models trained on scores simulated by GPT-4 with enhanced prompts.
\end{itemize}



\section{Results}

\subsection {\taskname\ using Scores on a Scale as Behavioral Information}

\begin{figure}[h]
  \centering
  \includegraphics[width=\linewidth]{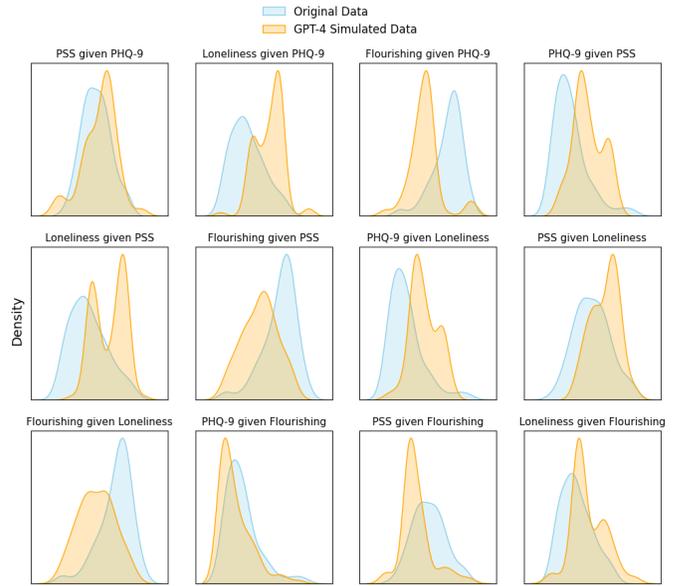}
  \caption{Density distribution of original and GPT-4 simulated scores for StudentLife dataset}
  \label{fig:density_plots}
\end{figure}

\begin{figure}[h]
  \centering
  \includegraphics[width=0.8\linewidth, height=0.4\linewidth]{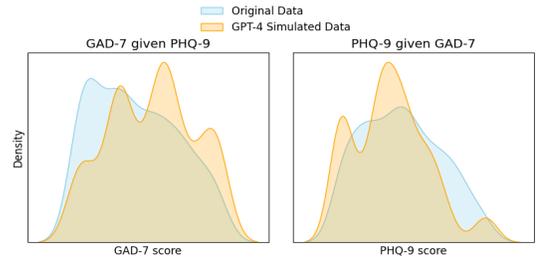}
  \caption{Density distribution of original and GPT-4 simulated scores for DepreST-CAT dataset}
  \label{fig:gad-phq-density}
\end{figure}

We evaluated the GPT-4 simulated scores using \taskname's evaluation metrics for StudentLife (Table \ref{tab:studentlife}) and DepreST-CAT (Table \ref{tab:deprestcat}) datasets. All the metrics in the tables are averaged over three runs of GPT-4 and have a low standard deviation ($\sigma <0.02)$ except when the Flourishing scale scores are used to simulate Loneliness scale scores  ($\sigma < 0.06$). We present the results across four metrics: 
\begin{itemize}
\setlength{\itemsep}{8pt}
      \item \textbf{Simulation Error}:
      As a measure of simulation error, we report the aggregated simulation error $(e_{aggregated})$.
      The $e_{aggregated}$ for different pairs of scales in the StudentLife dataset ranged from 15.3\% to 26.5\%, whereas the  $e_{aggregated}$ for the DepreST-CAT dataset were comparatively lower (12\%, and 13.6\% for $[\hat{R}_{GAD}]$ given $[{R}_{PHQ}]$, and $[\hat{R}_{PHQ}]$ given $[{R}_{GAD}]$, respectively). 
      
      \item \textbf{Correlation Alignment}: For the StudentLife dataset, the simulated participants' scores on a scale had low to moderate correlation with real scores on the scale. For the DepreST-CAT dataset, the simulated scores had a high correlation with real scores.


      \item \textbf{Distribution Alignment}: We used distribution density plots to visualize the distribution of simulated and real scores on different scales for the StudentLife (Figure \ref{fig:density_plots}) and DepreST-CAT (Figure \ref{fig:gad-phq-density}) datasets. We observed salient differences between the distributions.
      
    \item \textbf{Cross-Scale Validity:}
    We evaluated cross-scale validity between the scale scores used as behavior information and the scale scores being simulated.   
    \begin{itemize}
        \item \textbf{Trend Alignment}: For all scale pairs, GPT-4 simulated scores correctly aligned with the direction of correlation between scales as seen in real data.
    
        \item \textbf{Correlation Strength}: GPT-4 simulated scores consistently demonstrated higher strength of correlation between pairs of scales (except for ($[R_{Flourishing}]$, $[\hat{R}_{Loneliness}]$), resulting in a more monotonic relationship than is reflective in real data. 
\end{itemize}

\end{itemize}

To summarize, the performance of GPT-4 lies below expectations across several \taskname\ evaluation metrics, as GPT-4 establishes a stronger monotonic relationship between scales compared to real data. In DepreST-CAT dataset, however, simulating PHQ-9 scores using GAD-7 scores and vice-versa performs well in our evaluations. Therefore, we investigated the utility of using simulated PHQ-9 and GAD-7 scores in training ML models by replicating the original study on DepreST-CAT dataset.  


\begin{table*}[h]
\caption{Evaluation of \taskname\  application on StudentLife dataset }
\renewcommand{\arraystretch}{1.5}
\resizebox{\linewidth}{!}{%

\begin{tabular}{|c|cccccccccccccccc|}
\hline
                     & \multicolumn{16}{c|}{\textbf{Simulated Scale}}                                                                                                                                                                                                                                                                                                                                                                                                                                                                                                                      \\ \hline
                     & \multicolumn{4}{c|}{PHQ-9}                                                                                                                   & \multicolumn{4}{c|}{PSS}                                                                                                                     & \multicolumn{4}{c|}{Loneliness}                                                                                                             & \multicolumn{4}{c|}{Flourishing}                                                                                        \\ \hline
\textbf{Given Scale} & \multicolumn{1}{c|}{$M1$} & \multicolumn{1}{c|}{$M2$}    & \multicolumn{1}{c|}{$M3$} & \multicolumn{1}{c|}{$e_{agg}$}  & \multicolumn{1}{c|}{$M1$} & \multicolumn{1}{c|}{$M2$}    & \multicolumn{1}{c|}{$M3$} & \multicolumn{1}{c|}{$e_{agg}$}  & \multicolumn{1}{c|}{$M1$} & \multicolumn{1}{c|}{$M2$}   & \multicolumn{1}{c|}{$M3$} & \multicolumn{1}{c|}{$e_{agg}$}  & \multicolumn{1}{c|}{$M1$} & \multicolumn{1}{c|}{$M2$}    & \multicolumn{1}{c|}{$M3$} & $e_{agg}$  \\ \hline
PHQ-9                & \multicolumn{1}{c|}{\_}               & \multicolumn{1}{c|}{\_}       & \multicolumn{1}{c|}{\_}                & \multicolumn{1}{c|}{\_}     & \multicolumn{1}{c|}{0.53***}          & \multicolumn{1}{c|}{0.91***}  & \multicolumn{1}{c|}{0.52***}           & \multicolumn{1}{c|}{15.3\%} & \multicolumn{1}{c|}{0.33**}          & \multicolumn{1}{c|}{0.67***} & \multicolumn{1}{c|}{0.42***}           & \multicolumn{1}{c|}{19\%}   & \multicolumn{1}{c|}{-0.38***}         & \multicolumn{1}{c|}{-0.83***} & \multicolumn{1}{c|}{0.35**}            & 26.5\% \\ \hline
PSS                  & \multicolumn{1}{c|}{0.53***}          & \multicolumn{1}{c|}{0.79***}  & \multicolumn{1}{c|}{0.51***}           & \multicolumn{1}{c|}{21.2\%}   & \multicolumn{1}{c|}{\_}               & \multicolumn{1}{c|}{\_}       & \multicolumn{1}{c|}{\_}                & \multicolumn{1}{c|}{\_}     & \multicolumn{1}{c|}{0.32**}           & \multicolumn{1}{c|}{0.82***} & \multicolumn{1}{c|}{0.32**}            & \multicolumn{1}{c|}{17.4\%} & \multicolumn{1}{c|}{-0.52***}         & \multicolumn{1}{c|}{-0.92***} & \multicolumn{1}{c|}{0.52***}           & 21\%   \\ \hline
Loneliness           & \multicolumn{1}{c|}{0.33**}          & \multicolumn{1}{c|}{0.79***}  & \multicolumn{1}{c|}{0.34**}           & \multicolumn{1}{c|}{22\%}   & \multicolumn{1}{c|}{0.32**}           & \multicolumn{1}{c|}{0.74***}  & \multicolumn{1}{c|}{0.34**}           & \multicolumn{1}{c|}{17.5\%} & \multicolumn{1}{c|}{\_}               & \multicolumn{1}{c|}{\_}      & \multicolumn{1}{c|}{\_}                & \multicolumn{1}{c|}{\_}     & \multicolumn{1}{c|}{-0.56***}         & \multicolumn{1}{c|}{-0.91***} & \multicolumn{1}{c|}{0.5***}           & 19.2\% \\ \hline
Flourishing          & \multicolumn{1}{c|}{-0.38***}         & \multicolumn{1}{c|}{-0.84***} & \multicolumn{1}{c|}{0.42***}           & \multicolumn{1}{c|}{16.1\%} & \multicolumn{1}{c|}{-0.52***}         & \multicolumn{1}{c|}{-0.79***} & \multicolumn{1}{c|}{0.49***}           & \multicolumn{1}{c|}{18.3\%} & \multicolumn{1}{c|}{-0.56***}         & \multicolumn{1}{c|}{-0.25}   & \multicolumn{1}{c|}{0.14}              & \multicolumn{1}{c|}{15.4\%} & \multicolumn{1}{c|}{\_}               & \multicolumn{1}{c|}{\_}       & \multicolumn{1}{c|}{\_}                & \_     \\ \hline


\end{tabular}

}
\label{tab:studentlife}
\smallskip
\tiny

 $M1=corr([T_B], [T_A])$, $M2=corr([T_B], [\hat{T}_A])$, and $M3=corr([T_A], [\hat{T}_A])$ where $[T_A]$ and $[\hat{T}_A]$ are participants' aggregated real and simulated scores for the scale being simulated while $[T_B]$ is aggregated scores for the scale used as behavioral information and $corr(.)$ denotes Spearman's correlation test. Significance: ``no symbol'' $\rightarrow p > 0.05$, $*$ 
 $\rightarrow p \leq 0.05$, $**$ $\rightarrow p \leq 0.01$, and $***$ $\rightarrow p \leq 0.001$.
\end{table*}

\begin{table}[]
\caption{Evaluation of \taskname\ application on DepreST-CAT dataset }
\renewcommand{\arraystretch}{1.5}

\resizebox{\linewidth}{!}{%
\begin{tabular}{|c|cccccccc|}
\hline
                     & \multicolumn{8}{c|}{\textbf{Simulated Scale}}                                                                                                                                                                                  \\ \hline
                     & \multicolumn{4}{c|}{PHQ-9}                                                                                              & \multicolumn{4}{c|}{GAD-7}                                                                           \\ \hline
\textbf{Given Scale} & \multicolumn{1}{c|}{$M1$}    & \multicolumn{1}{c|}{$M2$}    & \multicolumn{1}{c|}{$M3$}   & \multicolumn{1}{c|}{$e_{agg}$}   & \multicolumn{1}{c|}{$M1$}    & \multicolumn{1}{c|}{$M2$}     & \multicolumn{1}{c|}{$M2$} & $e_{agg}$    \\ \hline
PHQ-9                & \multicolumn{1}{c|}{\_}      & \multicolumn{1}{c|}{\_}     & \multicolumn{1}{c|}{\_}     & \multicolumn{1}{c|}{\_}      & \multicolumn{1}{c|}{0.85***} & \multicolumn{1}{c|}{0.94***} & \multicolumn{1}{c|}{0.79***} & 12\% \\ \hline
GAD-7                & \multicolumn{1}{c|}{0.85***} & \multicolumn{1}{c|}{0.95**} & \multicolumn{1}{c|}{0.8***} & \multicolumn{1}{c|}{13.65\%} & \multicolumn{1}{c|}{\_}      & \multicolumn{1}{c|}{\_}      & \multicolumn{1}{c|}{\_}      & \_      \\ \hline

\end{tabular}
}
\smallskip
\label{tab:deprestcat}
\tiny

 $M1=corr([T_B], [T_A])$, $M2=corr([T_B], [\hat{T}_A])$, and $M3=corr([T_A], [\hat{T}_A])$ where $[T_A]$ and $[\hat{T}_A]$ are participants' aggregated real and simulated scores for the scale being simulated while $[T_B]$ is aggregated scores for the scale used as behavioral information and $corr(.)$ denotes Spearman's correlation test. Significance: ``no symbol'' $\rightarrow p > 0.05$, $*$ 
 $\rightarrow p \leq 0.05$, $**$ $\rightarrow p \leq 0.01$, and $***$ $\rightarrow p \leq 0.001$.

\end{table}

\subsection{DepreST-CAT study with \taskname\ simulated scores}

Tlachac et al., achieved desirable performance only for mild depression and anxiety screening (i.e., setting PHQ-9 and GAD-7 cutoff scores as 5) using two weeks of text and call data for crafting time series features \cite{tlachac2022deprest}. In our evaluations, we used the same features and show results for score cutoffs from 5 to 10.  We used simulated scores only for training and tested the performance on the held-out real data. 

\subsubsection{\textbf{Experiment B1}}

\begin{figure}[h]
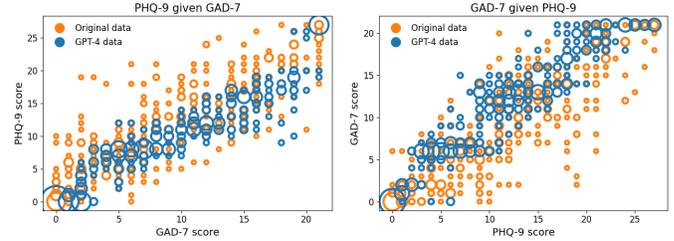

  \centering
  \begin{subfigure}[b]{0.49\linewidth}
    \includegraphics[width=\linewidth]{figures/scatter-phq.pdf}
    \label{fig:scatter-plot1}
  \end{subfigure}
  \begin{subfigure}[b]{0.49\linewidth}
    \includegraphics[width=\linewidth]{figures/scatter-gad.pdf} 
    \label{fig:scatter-plot2}
  \end{subfigure}
  \vspace{-8mm} 
    \caption{Scatter plots to compare the distribution of original scores and simulated scores when GAD-7 scores are used to simulate PHQ-9 scores (left) and vice-versa (right). Circle sizes are proportional to the number of data points. The GPT-4 simulated data is more linear making it less diverse.}
  \label{fig:scatter-plot}
\end{figure}

We first present scatter plots for real and GPT-4 simulated GAD-7 aggregated scores given participants' PHQ-9 scores and vice-versa (Figure \ref{fig:scatter-plot}). We observed that GPT-4 generates lower aggregated scores for PHQ-9 and higher aggregated scores for GAD-7. We hypothesize that since PHQ-9 questions are more negatively worded than GAD-7 questions and GPT models are aligned to be more positive, GPT-4 simulated lower scores on PHQ-9. We also observed that GPT-4 simulated scores demonstrated a stronger linear relationship between scores, making them less diverse compared to original scores.

\begin{figure}[h]
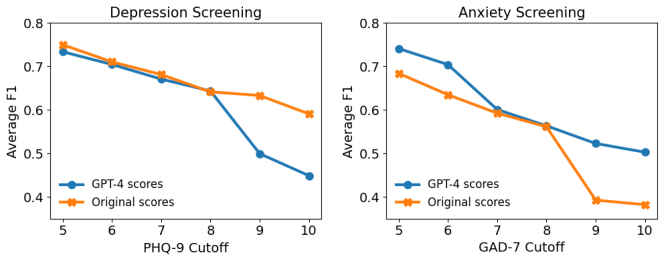

  \centering
  \begin{subfigure}[b]{0.49\linewidth}
    \includegraphics[width=\linewidth]{figures/xgboost_performance-dep.pdf}
    \label{fig:xgboost1}
  \end{subfigure}
  \begin{subfigure}[b]{0.49\linewidth}
    \includegraphics[width=\linewidth]{figures/xgboost_performance-anx.pdf} 
    \label{fig:xgboost2}
  \end{subfigure}
  \vspace{-8mm} 
  \caption{Performance comparison of F1-scores for XGBoost trained original and GPT-4 simulated scores for Depression (left) and Anxiety (right).}
  \label{fig:xgboost}
\end{figure}

We built the same five models as the original study, but for brevity, we present all our comparisons using XGBoost, which is the best-performing model, consistently attaining the highest F1 score. In Figure \ref{fig:xgboost}, we compare F1 scores for XGBoost on anxiety and depression screening tasks at different cutoffs. For anxiety screening, we observed that XGBoost trained on GPT-4 simulated scores performed better than XGBoost trained on original scores, whereas, for depression screening, XGBoost trained with GPT-4 simulated scores had a similar performance for cutoffs from 5 to 8 but gets worse as we go further. 


\subsubsection{\textbf{Experiment B2}}
\begin{figure}[h]
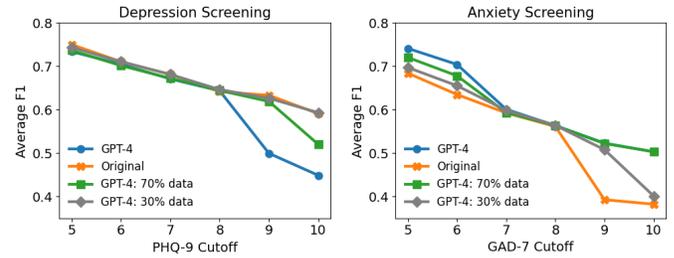

  \centering
  \begin{subfigure}[b]{0.49\linewidth}
    \includegraphics[width=\linewidth]{figures/ablation-dep.pdf}
    \label{fig:ablation1}
  \end{subfigure}
  \begin{subfigure}[b]{0.49\linewidth}
    \includegraphics[width=\linewidth]{figures/ablation-anx.pdf} 
    \label{fig:ablation2}
  \end{subfigure}
  \vspace{-8mm} 
   \caption{Performance comparison of XGBoost when 30\%, 70\%, and 100\% of training data is replaced with GPT-4 simulated scores.}
  \label{fig:ablation}

\end{figure}

In Figure \ref{fig:ablation}, we demonstrate that as we kept replacing real ground truth scores with GPT-4 simulated scores in training data, the performance gradually improved for anxiety screening models while gradually degraded for depression screening at higher cutoffs. 


\subsubsection{\textbf{Experiment B3}}
\begin{figure}[h]
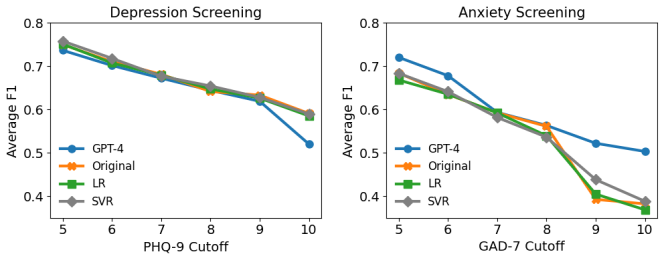

  \centering
  \begin{subfigure}[b]{0.49\linewidth}
    \includegraphics[width=\linewidth]{figures/linear-dep.pdf}
    \label{fig:linear-svr1}
  \end{subfigure}
  \begin{subfigure}[b]{0.49\linewidth}
    \includegraphics[width=\linewidth]{figures/linear-anx.pdf} 
    \label{fig:linear-svr2}
  \end{subfigure}
  \vspace{-8mm} 
   \caption{Performance comparison of XGBoost when trained on LR, SVR, and GPT-4 simulated scale scores.} 
  \label{fig:linear-svr}
\end{figure}

 In Figure \ref{fig:linear-svr}, we compare the performances of XGBoost, trained on the scores generated from LR, SVR, and GPT-4. LR and SVR exhibited similar performances to the original models, whereas GPT-4 scores consistently outperformed them in the anxiety screening task.


 \subsubsection{\textbf{Experiment B4}}
\begin{figure}[h]
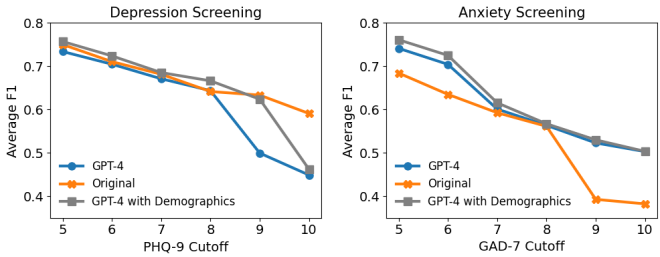

  \centering
  \begin{subfigure}[b]{0.49\linewidth}
    \includegraphics[width=\linewidth]{figures/demographics-dep.pdf}
    \label{fig:demographics1}
  \end{subfigure}
  \begin{subfigure}[b]{0.49\linewidth}
    \includegraphics[width=\linewidth]{figures/demographics-anx.pdf} 
    \label{fig:demographics2}
  \end{subfigure}
  \vspace{-8mm} 

   \caption{Performance comparison of XGBoost when trained on GPT-4 simulated scores with demographic information included in the prompt.}
  \label{fig:demographics}
\end{figure}

In Figure \ref{fig:demographics}, we demonstrate the performance comparison on using scores simulated by GPT-4 prompts with and without demographic information to train models. GPT-4 with demographics led to slightly improved performance in both anxiety and depression screening tasks, suggesting that adding additional context information in prompts could help GPT-4 develop better mental models of the participants.

\section{Discussion and Conclusion}

Digital mental health sensing studies have the potential for unobtrusive, automatic, and scalable screening of mental illnesses. Sensing technologies using supervised ML require ground truth labels for mental conditions obtained from participants' responses to multiple psychological scales. Responding to multiple scale items, however, can be burdensome for participants, especially in longitudinal studies with frequent and repeated measurements. To address this issue,  we proposed a novel framework (\taskname) to simulate participants' responses to a scale based on LLMs mental model of participants developed using information on their latent behavioral dispositions. We defined four evaluation metrics for assessing the alignment of simulated responses with real data. Additionally, we presented an application of the framework, using participants' responses to a different scale as the behavioral information provided to GPT-4 to simulate participants' responses on the desired scale. In our work, we used the GPT-4 simulated scores only to \textit{train} the ML models, following the approach by Tlachac~et~al.\ in their original study for depression and anxiety screening \cite{tlachac2022deprest}. It is important to note that we evaluated all models on held-out test data with the participants' real scores.


We discuss broader trends in LLM-simulated scale responses, methods to improve them, and their advantages and practical applications:


\subsection{Amplifying Monotonicity}
 Our results on the application of \taskname\ framework 
 demonstrated that GPT-4 is adept at gauging the direction of the correlation between simulated and provided scale scores while consistently amplifying the strength of the correlation, suggesting a more monotonic relationship than observed in real data. Consequently, GPT-4 tends to produce less diverse data, in line with previous research findings \cite{tavast2022language, hamalainen2023evaluating}.

 The models trained on these simulated scores may fail in cases where the scale scores are not monotonically related.  Our findings revealed that depression screening models trained on GPT-4 simulated PHQ-9 scores performed poorly on screening for moderate to high depression. We hypothesize that this discrepancy arises from the considerable variability in anxiety levels among individuals with moderate to high depression in our data (Figure \ref{fig:scatter-plot}). Conversely, anxiety screening models trained on GPT-4 simulated scores consistently outperformed the original models as the relationship between high anxiety and high depression is more monotonic in data, possibly making GPT-4 simulated labels act as noise-free data for training anxiety screening models and thus resulting in better performance. Our hypothesis is based on preliminary analysis of error rates in data and requires further investigation.

\subsection{Using Richer Behavioral Information}
We used participants' responses on a single psychological scale as behavioral information provided to LLMs. While our results are promising, we acknowledge that responses to a single scale may not fully capture participants' behavioral traits. It, however, serves as a good starting point. Adding additional demographic information boosted the performance of anxiety and depression screening models trained on simulated scores, highlighting a promising avenue for integrating richer contextual/behavioral information into prompts. Therefore, in future studies, we will explore prompting LLMs with richer behavioral sources like multiple scale responses, qualitative interviews, and medical histories.

\subsection{LLMs vs Conventional Approaches for Generating Scores}
Using LLMs to simulate responses on psychological scales offers substantial advantages over the conventional approach of building models from scratch. First, building models from scratch requires considerable time, effort, data, and domain expertise. LLMs, on the other hand, provide an easy-to-use and off-the-shelf solution requiring no additional data or training. Second, adding additional sources of information, such as demographics, might require modifications to model architecture and model retraining in conventional approaches while only requiring prompt engineering for LLMs. Third, LLMs can provide itemized scores and reasoning behind their scoring, whereas, implementing this feature in conventional approaches demands considerable effort.

\subsection{Practical Applications for Research Studies}
Simulating participants' responses on psychological scales has significant implications in addressing several challenges in mobile sensing and digital phenotyping studies. First, it can be employed for data imputation post-study to complete missing survey responses within datasets. Second, in longitudinal studies where it is burdensome for participants to complete multiple surveys daily, we can simulate some responses by building a mental model of them over time. It will be important to intermittently collect real responses to ensure the accuracy of the built mental model. Third, we can retrospectively analyze previously collected passive sensing datasets and develop mental health sensing models for outcomes for which researchers did not originally collect scale responses. Researchers often collect different subsets of surveys in longitudinal studies \cite{xu:GLOBEMDatasetMultiYear-2022,wang2014studentlife, mattingly2019tesserae,nepal:CapturingCollegeExperience-2024}, and \taskname\ could help reconstruct a common unified set of survey responses across datasets. We can achieve such reconstructions by building a mental model of participants using the data present in the dataset and then using LLMs to simulate their responses to new scales. Subsequently, we will be able to systematically benchmark models and methods to enable reproducible and repeatable findings~\cite{xu2023globem, mishra:stress-reproducibility, meegahapola2023generalization}. These applications serve as interesting future research directions.


\section*{Ethical Impact}

We used publicly available anonymized datasets and did not recruit any subjects for this work. The environmental impact of the energy used in prompting to GPT-4 API was negligible. As our proposed framework involves prompting LLMs with participants' behavioral information, we urge researchers adopting this framework to avoid sharing participants' Personally Identifiable Information (PII) with LLMs under any circumstances. Moreover, researchers planning to use participants' data to simulate their responses should inform participants about this use-case. The researchers should thoroughly evaluate the simulated responses to ensure that they do not misrepresent or underrepresent any groups. As we evaluated the utility of simulated scores in training ML models using a single dataset, further research is necessary to claim the generalizability of our results in different scenarios. Furthermore, due to the critical nature of mental health sensing and the potential ramifications of changes in ML model performances, we discourage wholly substituting real human data in constructing mental health sensing models. Instead, we envision this approach to complement naturalistic and observational human data. We insist on rigorous testing of models trained on simulated data before their deployment. 

\section*{Acknowledgment}

This work was partially supported through compute resources provided by Microsoft as a part of their AICE (AI, Cognition, and the Economy) projects and by the NIH National Institute of Drug Abuse under award number NIH/NIDA P30DA029926. The views and conclusions contained herein are those of the authors and should not be interpreted as necessarily representing the official policies, either expressed or implied, of the sponsors. Any mention of specific companies or products does not imply any endorsement by the authors, by their employers, or by the sponsors






\balance

\bibliographystyle{ieeetr}
\bibliography{references}


\end{document}